\documentstyle{laa}

\def\be{\begin{equation}}
\def\ee{\end{equation}}
\def\bea{\begin{eqnarray}}
\def\eea{\end{eqnarray}}
\def\eqref#1{\ref{eq#1}}
\def\e#1{\label{eq#1}}
\def\Rt{{\tilde R}}

\def\intsemi{\int_{0}^{\infty}}
\def\intall{\int_{-\infty}^{\infty}}
\def\phminus{\phantom{-}}
\def\nf{P}  % number of frequencies
\def\Teff{T_{\rm eff}}
\def\hi{H\,{\sc i}}

\def\hei{He\,{\sc i}}
\def\heii{He\,{\sc ii}}
\def\hneut{\mbox{H$^0$}}
\def\heneut{\mbox{He$^0$}}
\def\heplus{\mbox{He$^+$}}
\def\feiv{Fe\,{\sc iv}}
\def\fev{Fe\,{\sc v}}

\begin{document}
\thesaurus{03(02.12.1,02.18.7,02.19.2,08.01.3,08.05.1)}

\title{An accelerated lambda iteration method for multilevel radiative transfer}
\subtitle{III. Noncoherent electron scattering}
\author{G.B. Rybicki \inst{1}
\and D.G. Hummer \inst{2,} \inst{3}}
\institute{Harvard-Smithsonian Center for Astrophysics, 60 Garden Street,
  Cambridge, MA 02138 USA
\and Max-Planck-Institut f\"ur Astrophysik, Postfach 1523, D-85740 Garching, 
  Germany
\and Institut f\"ur Astronomie and Astrophysik der Universit\"at M\"unchen,
  Scheinerstr.\ 1, D-81679 M\"unchen, Germany}

\date{Received date; accepted date}

\maketitle

\begin{abstract}  

Since the mass of the electron is very small relative to atomic masses, Thomson 
scattering of low-energy photons ($h\nu \ll m_ec^2$) produces thermal
Doppler frequency shifts that are much
larger than atomic Doppler widths. A method is developed here to evaluate the
electron scattering emissivity from a given radiation field which 
is considerably faster than previous methods based on 
straightforward evaluation of the scattering integral. This procedure is
implemented in our multilevel radiative code (MALI), which now takes 
full account of the effects of noncoherent electron
scattering on level populations, as well as on the emergent spectrum.
Calculations using model atmospheres of
hot, low-gravity stars display not only the expected broad wings of strong
emission lines but also effects arising from the scattering of photons across
continuum edges. In extreme cases this leads to significant shifts of the 
ionization equilibrium of helium.

\keywords{radiative transfer -- noncoherent Thomson scattering -- non-LTE -- 
  stellar atmospheres}
\end{abstract}

\section{Introduction}

The scattering by electrons of low-energy photons ($h\nu \ll m_ec^2$) is 
important in astrophysical situations only under the rather special 
circumstance that the electron-scattering opacity is not too much smaller
than that from all other sources.  Thomson scattering in these cases has
traditionally been regarded as coherent in frequency, and much work on
spectrum formation has been based on this assumption.
However, if the spectrum of the object in questions contains features with
a width less than or comparable to the electron Doppler width, such as strong
lines and continuum edges, the scattering may have to 
be treated as noncoherent, i.e. the change in the frequency of the scattered
photon by the electron Doppler effect must be taken into account (M\"unch 
\cite {GM}, Hummer \& Mihalas 
\cite{HM}, Auer \& Mihalas \cite{AM1}, \cite{AM2}).  
Although the description of this 
redistribution process as a convolution of the mean intensity $J(\nu)$ with
a redistribution function is simple in 
principle, its inclusion in numerical calculations leads to considerable 
difficulties in practice for three reasons. 1) The scale of the redistribution 
in frequency is much larger than the atomic Doppler line widths; 2) the 
evaluation of the convolution at each frequency involves an integration over a 
wide band of surrounding frequencies; 3) this coupling of frequencies conflicts 
with the basic strategy of approximate lambda iteration (ALI)
methods which involve a frequency-by-frequency 
evaluation of the approximate lambda operator.

The transfer equation for two level systems accounting for noncoherent electron
scattering 
have been solved in various approximation by a number of workers. Early work 
on this problem was based on the formulation in terms of a ``reversing
layer'' by Chandrasekar (\cite{C48}), who found a solution for the case in which
the reddening of scattered photons arising from the Compton effect was
 included. Subsequently
M\"unch (\cite{GM}) considered the effects of electron
Doppler redistribution, again
in terms of a ``reversing-layer'' model, by means of a Fourier transform in
frequency.
This and other early work is
summarized in Chapter 12 of Chandrasekhar (\cite{RT}).
Numerical solutions have been given for lines in the atmospheres of O-type stars
by Auer and Mihalas (\cite{AM1}, \cite{AM2}),who assumed complete redistribution
for the atomic scattering. Rangarajan et al (\cite {RMP}) give solutions for
parameterized models with both both partial and complete redistribution in
the line. 

Hillier (\cite{Hi}) and Hamann et al.\ (\cite{Hamann})
have included noncoherent electron scattering (NES) 
in computing the emergent line spectra of
realistic models of Wolf-Rayet stars, for which the effects are clearly
observable.  However, in these works NES is
included only during the formal solution for the emergent spectra, 
after the level populations have
been fixed from an NLTE solution in which NES 
has not been taken into account.

In this paper we develop a numerical method for 
the solution to radiative transfer problems in which
noncoherent electron scattering can play an important role.
The method is based on approximating the electron scattering
redistribution function by a sum of exponentials.
It is shown that only two exponential terms give a very accurate approximation.
The electron scattering 
emissivity as a function of frequency at each point in the atmosphere then 
can be expressed through the solution to two simple differential equations in 
frequency space. The
corresponding difference equations can be solved numerically by Gaussian 
elimination.  This procedure is easily included in any iterative solution of
the combined radiative transfer and statistical equilibrium equations. 

In previous schemes for evaluating the NES emissivity, the computing
time for each depth in the atmosphere 
scales as $N_FN_W$, where $N_F$ is the number of frequency grid points,
and $N_W$ is the number of frequency grid points needed to 
represent adequately the width of the
electron redistribution function.  For applications to spectral
lines with structure on scales of both atomic and electron Doppler widths, 
$N_W$ can be of order of 10--50 or larger.
The present method scales more favorably as 
$CN_F$, where $C$ is a small constant, independent of the frequency
grid. It also automatically enforces exact conservation of photon number.

In this paper we describe an implementation of our method of 
treating noncoherent electron
scattering for the multilevel ALI code MALI 
(Rybicki \& Hummer, \cite{RHI}, \cite{RHII}; hereafter RHI, RHII).
In contrast to previous methods (Hillier \cite{Hi}, Hamann et al.
\cite{Hamann}), full account is taken of 
the effect of NES on the level populations,
as well as on the emergent spectrum.
Although the particular application described here is to MALI,
the method presented here is quite general and should be easily 
adaptable to other codes as well, especially ALI codes.

It should be made clear that in this latest generalization of MALI 
noncoherent electron scattering is treated not fully by ALI, but only 
by means of {\em ordinary} lambda iteration.  This was not by choice,
but resulted from a failure
to find suitable approximate electron scattering
operators to use in an ALI method.
In particular, we tried several approximate operators
based on coherent scattering, which 
is already solved non-iteratively in MALI.  
All of our choices were either unstable or had very poor convergence
properties.  

Fortunately, for many cases of interest, the mean number of 
photons scatterings due to electron scattering is very moderate,
of order a few tens. Since the typical number of iterations in
an ALI solution can be of order many tens to a hundred, the treatment of
NES by ordinary lambda iteration in these cases will not substantially change
the net number of iterations required for a solution.
One should also note that numerical accelerators, 
such as Ng's (\cite{Ng}) method, used in MALI, act to
improve the convergence of ordinary lambda iteration, as well as ALI.
However, for problems with mean numbers of scatterings
of order of a hundred or more, 
the present method is simply not suitable,
and other methods will have to be
developed.

In Sect.\ 2 we present the basic description of the electron redistribution
problem. In Sect.\ 3 we develop the approximate exponential fit to
the electron scattering redistribution function and show how the
emissivity can be found by solving two differential equations.
We then give results 
illustrating the accuracy of our approximation and the basic features of 
two-level transfer problems including electron scattering treated as coherent 
and incoherent processes.  Sect.\ 4 describes the incorporation
of our method into MALI and gives results for a hot O-star atmosphere
including noncoherent electron scattering.
As expected, the effects on individual lines were quite noticeable,
as were the effects within a few electron Doppler widths of continua.
However, we also found an unexpected 
and potentially important effect for certain continua,
which, in the absence of noncoherent scattering, are strongly in absorption.
Then scattering of radiation from
the stronger continuum below the edge into the other can cause a substantial
anomalous ionization of material,
and conseqently a substantial change in the radiation field
across the entire ionizing continuum, not just in the neighborhood of the jump.

\section{Basic equations}

     The angle-averaged emissivity at frequency $\nu$ for 
unpolarized electron scattering 
in the non-relativistic limit can be written
$j(\nu)=\sigma_T E(\nu)$, where $\sigma_T$ is the Thomson cross
section and $E(\nu)$ is the scattering integral,
\be
  E(\nu) \equiv \intsemi R(\nu,\nu')J(\nu')\,d\nu'.  \e{2.1}
\ee
Here $J(\nu)$ is the mean intensity, and $R(\nu,\nu')$ is the unpolarized,
angle-averaged electron scattering redistribution function.  
This relation holds at each spatial point in the medium; however, we
shall suppresss this spatial dependence in the notation.
The {\em noncoherent} scattering expressed by Eq.\ (\eqref{2.1}) 
is to be distinguished from {\em coherent} scattering,
for which $E(\nu)=J(\nu)$.

Hummer \& Mihalas (\cite{HM}) derived explicit forms for the 
redistribution functions for non-relativistic electron scattering 
assuming negligible Compton energy shift.  
It is convenient for our purposes to express their results
in terms of a modified function $\Rt(y)$, 
\be
   R(\nu,\nu')={1\over \beta_T\nu'}\Rt(y),   \e{2.2}
\ee
where the variable $y$ is defined by
\be
  y= {\ln \nu-\ln \nu'\over \beta_T} 
     = {1\over \beta_T}\ln {\nu\over \nu'}. \e{2.3}
\ee
The quantity $\beta_T$ is the mean electron thermal speed divided by
the speed of light $c$,
\be
 \beta_T = {1\over c}\sqrt{{2kT\over m_e}}=1.84\times 10^{-5}\, T^{1/2}, \e{2.4}
\ee
where $T$ is the temperature in Kelvin, $m_e$ is the electron mass, and
$k$ is Boltzmann's constant.

Hummer \& Mihalas (\cite{HM}) gave two forms
for $\Rt$, depending on whether isotropic scattering or
the more exact dipole scattering is assumed; these are usually distinguished
by the subscripts $A$ and $B$, respectively.  
Their results are\footnote{The occurrences of ``erf'' in the formulas of
Hummer \& Mihalas (\cite{HM}) are typographical errors, which
should be replaced by ``erfc''.}:
\bea
   \Rt_A(y) &=& \hbox{ ierfc }{|y|\over 2} 
             ={1 \over \sqrt{\pi}}e^{-y^2/4} 
    - {|y| \over 2} \hbox{ erfc } {|y| \over 2}, \e{2.5} \\
    \Rt_B(y) &=&{3\over 2}\hbox{ ierfc}
     {|y| \over 2}  -12\hbox{ i$^3$erfc}{|y| \over 2} 
     +96\hbox{ i$^5$erfc}{|y| \over 2} \nonumber\\
  &=& ({11\over 10}+{2\over 5}y^2+{1\over 20}y^4)
 {1\over \sqrt{\pi}}e^{-y^2/4} \nonumber \\
    &&{\hskip 0.5cm} - ({3\over 2}+{1\over 2}y^2+{1 \over 20}y^4)
       {|y| \over 2}\hbox{ erfc } {|y| \over 2}. \e{2.6}
\eea
The functions $\hbox{ i$^n$erfc}$ denote repeated integrals of the
error function, defined, e.g., in Abramowitz \& Stegun 
(\cite{AS}; Sect.\ 7.2).  These have been expressed in terms of the ordinary
error function erfc by means of recurrence relations.  

Formulas (\eqref{2.5}) and (\eqref{2.6}), when substituted into
Eq.\ (\eqref{2.2}), give redistribution functions that
differ from those given by Hummer \& Mihalas (\cite{HM}),
in that the ``line center'' frequency $\nu_0$ does not appear, and the variable 
$y$ is now defined in terms of logarithms of the frequencies $\nu$ and $\nu'$.
These changes have been made so that the resulting formulas apply
for the whole spectrum, not just to a region in the neighborhood of
a single line.  It is appropriate in the non-relativistic limit
that the redistribution process should depend primarily 
on the {\it ratio} of frequencies rather than the
{\it differences} that appear in the Hummer and Mihalas formulas.
This dependence on frequency ratio comes about because  
the Compton energy shift is negligible in the non-relativistic limit
and the energy shift of a photon in the scattering process is due solely to the
Doppler shifts into and out of the rest frame of the electron.
(see, e.g., Rybicki \& Lightman \cite{RL}; Sect.\ 7.3).

It can easily be verified that formulas (\eqref{2.5}) and (\eqref{2.6}) 
lead to the
Hummer \& Mihalas forms when applied to line transfer for
cases of interest in stellar atmospheres,
where $\beta_T \ll 1$, implying that the
redistribution functions are sharply peaked 
near $\nu \approx \nu'$.
An appropriate expansion of $y$ is then
\be
  y \approx {\nu -\nu' \over \beta_T\nu_0},  \e{2.7}
\ee
where $\nu_0$ can be taken to be either $\nu$ or $\nu'$, or, as
in Hummer \& Mihalas (\cite{HM}), the line center frequency $\nu_0$.
Likewise, the overall factor in Eq.\ (\eqref{2.2}) can be replaced by
$1/(\beta_T \nu_0)$.

However, for the present work the full logarithmic
definition (\eqref{2.3}) for $y$ will be used.  In fact, it is convenient to
introduce the logarithmic frequency variable
\be
  \xi = \log \nu,  \e{2.8}
\ee
which will be used instead of the frequency $\nu$.  Making the
change of variables in Eqs.\ (\eqref{2.1}) and (\eqref{2.2}), we
obtain 
\be
  E(\xi) = \beta_T^{-1} \intall \Rt( \beta_T^{-1}|\xi - \xi'|) J(\xi')\,d\xi',
   \e{2.9}
\ee
where $E(\xi)$ and $J(\xi)$ now denote the scattering integral
 and mean intensity as functions of the variable $\xi$.

Since photon numbers are conserved upon scattering, the following integrals
should be equal
\be
 \intsemi {E(\nu)\, d\nu \over h\nu} 
        = \intsemi {J(\nu)\, d\nu \over h\nu}. \e{2.10}
\ee
(The intensities here are defined in terms of energy, so one must
divide by $h\nu$ to convert to photon numbers.)
In terms of the logarithmic variable $\xi$, this relation may be
written,
\be
     \intall E(\xi)\,d\xi = \intall J(\xi)\,d\xi. \e{2.11}
\ee
Substituting Eq.\ (\eqref{2.9}) and demanding that the resulting
equation hold for all functions $J(\xi)$,  we find the normalization
condition
\be 
   \intall \Rt(y)\,dy = 1,  \e{2.12}
\ee
which is exactly satisfied for both forms (\eqref{2.5})
and (\eqref{2.6}).  

This normalization condition (\eqref{2.12}) also justifies
the use of coherent scattering as an approximation 
when the scale of variation of the
radiation field is large compared to the electron
Doppler width.  In that case, $J(\xi')$ can be taken from
under the integration in Eq.\ (\eqref{2.9}), replacing $\xi'$
by $\xi$, and the normalization condition implies the 
coherent result $E(\xi)=J(\xi)$.  Coherent scattering
is usually a good approximation for continua, but it can fail badly 
in the neighborhood of lines and continuum edges, as we shall see.

Another quantity of importance is the second moment
\be 
  \intall \Rt(y) y^2\,dy = {1\over 2},  \e{2.13}
\ee
which measures the effective width of the redistribution function.
This moment has the same value ($1/2$) for both forms (\eqref{2.5})
and (\eqref{2.6}).
The second moment is particularly
important for describing noncoherent electron scattering
in cases where large numbers of scatterings occur,
such as in the far wings of lines.
In this case, the frequency behavior can be well described
by a random walk 
that depends solely on the two moments (\eqref{2.12}) and  (\eqref{2.13}).

By contrast, the fine details in the centers of lines depend primarily
on the absolute value of the function $\Rt(y)$ near $y=0$.  For
this reason the value of $\Rt(0)$
is very important, and also, to a lesser extent, its (right) derivative
$\Rt'(0^+)$.
These quantities differ for the isotropic and
dipole cases, and are given by,
\be
\Rt_A(0)={1\over \sqrt{\pi}}, \qquad   \Rt_A'(0^+)=-{1\over 2}, \e{2.2.2a}
\ee
\be
 \Rt_B(0)={11\over 10\sqrt{\pi}}, \qquad \Rt_B'(0^+)=-{3\over 4}. \e{2.2.2b}
\ee

\section{The exponential approximation}

An efficient method for the evaluation of the integral in Eq.\ (\eqref{2.9})
can be based on a simple
approximation to the modified redistribution functions $\Rt(y)$
in the form of a sum of $N$ exponential terms:
\be
 \Rt(y) = {1\over 2}\sum_{i=1}^N a_ib_i \exp(-b_i |y|).   \e{2.2.1}
\ee
We shall call this the
{\em exponential approximation}, but
it should be noted that, 
in terms of true frequencies, the approximation actually takes the 
{\em power law} form,
\be
 \Rt(y) = {1\over 2}\sum_{i=1}^N a_ib_i
         \left( {\nu_< \over \nu_>} \right)^{b_i/\beta_T},   \e{2.2.1b}
\ee
where $\nu_<$ and $\nu_>$ are, respectively, the smaller and larger 
of $\nu$ and $\nu'$. 

\subsection{Fitting procedures}

Sets of constants $a_i$ and $b_i$ can be determined in a number of ways,
all of which depend on matching some properties of the approximate
and true functions.  In view of the importance of conditions 
(\eqref{2.12}) and (\eqref{2.13}), we demand that coefficients should 
accurately satisfy the relations
\be
  \sum_{i=1}^N a_i = 1,     \e{2.2.1c}
\ee
\be
  \sum_{i=1}^N a_i b_i^{-2} = {1 \over 2}.  \e{2.2.1d}
\ee

The simplest exponential approximation consists of one term ($N=1$),
with two independent coefficients $a_1$ and $b_1$ chosen to satisfy both
Eqs.\ (\eqref{2.2.1c}) and (\eqref{2.2.1d}), namely,
$a_1=1$ and $b_1=\sqrt{2}$.
Since Eqs.\ (\eqref{2.2.1c}) and (\eqref{2.2.1d}) are the same for
both isotropic and dipole forms for $\Rt(y)$, 
this approximation does not distinguish between them.

Improved approximations can be obtained by using two terms ($N=2$),
with four independent coefficients, giving two additional degrees 
of freedom as compared with the single term approximation.
In view of the importance of the behavior of $\Rt(y)$ near $y=0$,
we would like to use these new degrees of freedom to fit the
value and slope of the
redistribution function at the origin, given by Eq.\
(\eqref{2.2.2a}) or (\eqref{2.2.2b}).  These conditions
require that the following two equations be satisfied,
\be
\sum_{i=1}^N a_ib_i = 2\Rt(0),  \e{2.2.1e}
\ee
\be
\sum_{i=1}^N a_ib_i^2 = -2\Rt'(0^+),  \e{2.2.1f} 
\ee

Note that the four conditions expressed by
Eqs.\ (\eqref{2.2.1c}) through (\eqref{2.2.1f})
are all of the same general form, 
\be
\sum_{i=1}^N a_ib_i^{m} = S_m,  \e{2.2.1g}
\ee
with different values of $m$.  Approximations for larger
$N$ constructed by fitting higher order moments and
higher order derivatives will also satisfy equations of this
form.

Equation (\eqref{2.2.1g}) represents a set of nonlinear
equations for the unknown values of the coefficients $a_i$ and $b_i$.
In general, numerical methods must be used to solve 
them\footnote{In the special case where the values of $m$ are 
equally spaced, equations of the form (\eqref{2.2.1g}) can be
solved analytically by Prony's method (see, e.g., Hildebrand \cite{Hilde}).
However, Prony's method often produces complex solutions.}.
We used the Newton-Raphson iterative method and confined our search to
real solutions; complex solutions imply oscillating exponential terms that
give unacceptable negative values for $\Rt(y)$ for large $y$.

We attemped to solve the above $N=2$ system of equations for the
isotropic and dipole functions.  However, these nonlinear equations
apparently do not have (real) solutions when both the value and slope 
are given their exact values.  Thus we chose to fit the value exactly
and to fit the slope as closely as possible,
namely, -0.333 (instead of -0.5) for the isotropic case and -0.6
(instead of -0.75) for the dipole case.
The results obtained in this way are summarized in Table 1.

\begin{figure}
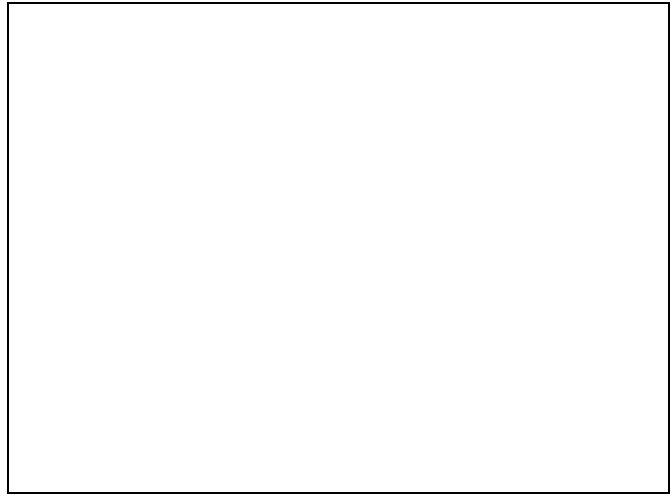

\picplace{6.5 cm}
\caption[]{Exact and approximate redistribution functions 
  $\Rt(y)$ for a) isotropic and b) dipole}
\end{figure}

\begin{table}
\caption[]{Coefficients for exponential approximation}
\begin{flushleft}
\begin{tabular}{lllll}
\hline
Type & $N$ & $i$ & $a_i$ & $b_i$ \\
\hline
A1\&B1 & 1 & 1 &   \phminus 1.000000000000 &  \phminus 1.414213562373 \\
\hline
A2 &   2 & 1  &  \phminus 5.682790496584 &  \phminus 1.835757192503 \\
  &     & 2  &          -4.682790496584 &  \phminus 1.986816272751 \\
\hline
B2  &  2 & 1  &  \phminus 1.690703717290 &  \phminus 1.614249968779 \\
   &    & 2  &          -0.690703717290 &  \phminus 2.154326524957 \\
\hline
\end{tabular}
\end{flushleft}
\end{table}

In Fig.\ 1 the exact functions $\Rt$ for isotropic and dipole case
are plotted along with the corresponding 
exponential approximations.  The maximum relative errors over the range
$0 \le y \le 3$ for the $N=1$ approximations are about 25\% for the
isotropic and about 17\% for the dipole.  
The maximum relative errors for $N=2$ approximations over this range are
both about 10\%.
For values beyond $y=3$ the relative error 
of the approximation increases rapidly, since the true redistribution
function decays asymptotically faster than exponential.  
However, at $y=3$ the absolute values of these functions 
have already decreased by about two orders of magnitude from their
values at $y=0$, so the region $y > 3$ should not be a substantial
source of error in the emissivities.

There is reason to expect solutions based on these
approximations to be more accurate
than the above maximum errors might suggest.  Note that
these approximations have been specifically constructed  to be
exact for the values and moments of the redistribution functions
that are most critical for the solution.  Also, the treatment of
electron scattering involves integration over the redistribution
function and this tends to average out the errors. 

\subsection{Reduction to differential equations}

One of the advantages of the exponential representation is that the evaluation
of the scattering integral $E$ can be reduced to the solution of 
differential equations.
To see this, let us substitute Eq.\ (\eqref{2.2.1}) 
into Eq.\ (\eqref{2.9}), leading to
\be
   E(\xi) = \sum_{i=1}^N a_i F^{(i)}(\xi), \e{2.2.2.1}
\ee
where
\be
   F^{(i)}(\xi) = {1\over 2}b_i\beta_T^{-1}\intall 
         \exp(-b_i\beta_T^{-1}|\xi-\xi'|) J(\xi')\, d\xi'.  \e{2.2.2.3}
\ee
Differentiating twice with respect to $\xi$, one easily shows that
$F^{(i)}(\xi)$ satisfies the second-order differential equation
\be
-{\beta_T^2 \over b_i^2} {d^2F^{(i)}(\xi) \over d\xi^2} +F^{(i)}(\xi) = J(\xi). 
        \e{2.2.2.4}
\ee

One property of this differential approximation is that it exactly
preserves the normalization condition (\eqref{2.11}).  To show this,
we integrate Eq.\ (\eqref{2.2.2.4}) over $\xi$, which gives
\be
   \intall F^{(i)} (\xi)\,d\xi = \intall J(\xi)\,d\xi, \e{2.2.2.5}
\ee
assuming that the first derivative of $F^{(i)}$ vanishes at both
endpoints of the region of integration.  
This will generally be true if the frequency range is
chosen so that the radiation field is 
sufficiently small at the limits of the range, or by explicit
choice of boundary conditions (see Appendix A).
Now multiplying by $a_i$ and summing over all $i$, using Eqs.\ 
(\eqref{2.2.1c}) and (\eqref{2.2.2.1}),
the normalization condition (\eqref{2.11}) is recovered.

The radiation transfer problem can be treated numerically
by introducing a frequency grid $\xi_j$, $j=1,\ldots,\nf$.  
All functions of frequency are then represented by the set of
their values on this grid, e.g., $E_j = E(\xi_j)$.
Eq.\ (\eqref{2.2.2.1}) then gives the values of electron scattering
emissivity on the grid,
\be
    E_j = \sum_{i=1}^{N} a_i  F^{(i)}_j, \e{2.2.3.5}
\ee
in terms of the quantities $F^{(i)}_j = F^{(i)}(\xi_j)$.
These quantities can be determined by solving the differential
equation (\eqref{2.2.2.4}) for each value of $i$ by means of the 
well known Feautrier method.  The details of the numerical
method are given in Appendix A.  A proof that the numerical method 
also preserves
the correct normalization (\eqref{2.11}) is given in Appendix B.

\subsection{Relation to the Kompaneets equation}

The differential equation method of the last section bears some superficial
similarity with the treatment of electron scattering using 
the Kompaneets equation (Kompaneets \cite{Komp}; Rybicki and Lightman \cite{RL},
Sect.\ 7.6), which is closely related to the Fokker-Planck equation.
However, as we now explain, the two methods are very different 
in their range of validity and in their mathematical structure.

The major advantage of the 
Kompaneets equation is that it includes the major physical effects
associated with electron scattering:
1) the frequency spreading due to the thermal Doppler effect;
2) the frequency shift due to thermal Doppler effect (inverse Compton effect);
3) the frequency shift due to electron recoil (Compton effect); 
and 4) stimulated electron scattering.  
Because of this, it is applicable to many aspects of Comptonization phenomena.
Its major limitation
is that it can only describe radiation fields
that vary slowly on the scale of the frequency shift per
scattering.  In particular, the Kompaneets equation is not
appropriate for treating the neighborhoods of lines or
continuum edges.

The differential equation method of this paper treats only the
first physical effect above, namely, the thermal Doppler spreading effect.
However, its major advantage is that
it can treat radiation
fields that vary rapidly on the scale of typical frequency shifts.
This is essential for treating lines and continuum edges 
in typical stellar atmospheres, where, fortunately, the
other physical effects treated by the Kompaneets equation
are usually not very important.

The mathematical distinction between the two methods is 
very striking. Let us compare the Kompaneets equation 
with the $N=1$ approximation of this paper, where $E(\xi)=F^{(1)}(\xi)$.  Then
both methods are formulated using second-order differential
operators.  However, the Kompaneets equation expresses the
{\em emissivity} as a differential operator acting on the
{\em radiation field}; in contrast, the differential equation
(\eqref{2.2.2.4}) expresses the {\em radiation field} as a
differential operator acting on the {\em emissivity}.
Thus, our second-order operator is not
directly comparable to the Kompaneets operator
but rather to its {\em inverse}.
It is this distinction that allows
our method to treat rapidly varying radiation fields.

A further mathematical point is that our method can be
improved by going to larger values of $N$, with no noticeable
numerical problems.  An analogous
improvement of the Kompaneets equation would involve the introduction of
higher order differential operators; however, it is known that
this can lead to numerical problems, such as instabilities and
negative solutions.

\subsection{Tests of the method}

In order to test the exponential approximation method, 
it was applied to some simple, parametrized models.
A specially written, non-iterative numerical code was developed for
this purpose,
so that the results could be evaluated directly, 
without being confused by questions of iterative convergence.
We shall present one particular such parametrized model here;
another will be presented in Sect.\ 4.2.

\begin{figure}
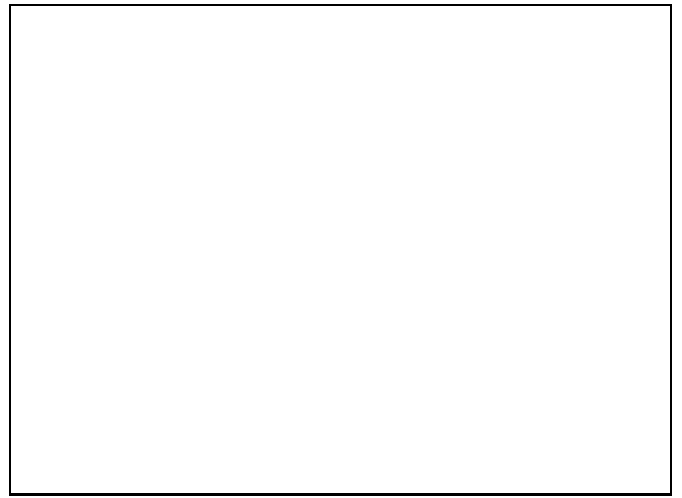

\picplace{6.5 cm}
\caption[]{$\log F_x$ (arbitrary units) vs. $x$  
for parametrized model using 
  coherent and various noncoherent scattering approximations}
\end{figure}

In Fig.\ 2 is shown one result of applying the non-iterative
code.  
The case is one of pure line scattering in 
a finite slab of total line mean optical depth $\tau_{\rm L}=2\times 10^4$ 
and total electron scattering optical depth $\tau_{\rm e}=20$.  
The photon injection rate into the line is uniform with depth.
The log of the emergent flux $F_x$ (in arbitrary units) 
is plotted against $x$, the frequency
relative to line center in units of the line Dopper width.
The mass of the ion is assumed to be that of hydrogen, so that the
electron Doppler width is $\sim 43$ in these units.
This is an extreme case for which line photons are scattered many times 
($\sim \tau_{\rm e}^2$) by the electrons, 
leading to wide wings extending out to many line Doppler widths.
The two parts of this figure
examine the flux over very different frequency bands:  Fig.\ 2a
shows the details near the center of the line, $|x| < 20$, while Fig.\ 2b
shows the behavior in the far electron-scattering wings, $|x|<1000$.

Four types of curves are plotted in Fig.\ 2.  The curves marked C give the
result assuming that the electron scattering is coherent.
The remaining curves are marked according to the type of
redistribution function assumed, A for isotropic and B for dipole,
and also by the number of terms in the exponential approximation.

One sees immediately that the coherent solution is vastly different from the
other three solutions, which are themselves very similar.
The behavior for large $x$ shown in Fig.\ 2b
is particularly striking, in that all the
noncoherent solutions are graphically indistinguishable in the
electron-scattering wings; this is explained
simply as the result of all three approximations having the correct
zeroth and second moments of the redistribution function.  
The only place where distinctions between
the three noncoherent cases 
can be seen is just outside the line core, where the differences
between the values of $\Rt(0)$ are important.

The overwhelming impression from Fig.\ 2 is that the differences in the
solutions due to different forms of the noncoherent approximation
are far less important than the differences with the coherent approximation.
We conclude that any of the various noncoherent
approximations will do acceptably well.
{}From a theoretical point of view we prefer the dipole form, since it
is physically more realistic than the isotropic one.  Therefore,
all further numerical calculations reported in this paper
will use the $N=2$ dipole approximation (B2).

\section{Implementation in MALI}

The form of the approximate lambda iteration method developed in RHI and RHII is
implemented in the FORTRAN program MALI. Since the publication of RHII the
code has been generalized to treat line overlaps and an arbitrary number
of chemical species.

Although the implementation described here is specific for MALI, it
should be easily adaptable to other numerical codes, especially
to ALI codes.

\subsection{Electron scattering by lambda iteration}

The evaluation of the electron-scattering emissivity described
in Sect.\ 3 has been implemented in MALI to treat noncoherent electron 
scattering by lambda iteration, while all bound-bound and bound-free 
processes continue to be treated with our ALI procedure.

The frequency grid is constructed with double points at every continuum edge
to allow for discontuities in the radiation field. On both sides of every edge
and line up to twenty additional frequency points are included  with a spacing
of a specified fraction of the electron scattering Doppler width at some
fiducial temperature. A mapping is constructed to and from a secondary 
frequency grid without double points, on which the difference Eqs.\
(\eqref{2.2.3.1}) with boundary conditions (\eqref{2.2.3.6}) are to
be integrated, and the factors weighting $J^-_j$ and $J^+_j$ in Eq.\
(\eqref{2.2.5.1}) are
stored. In each iteration, after $J_{\nu}$ is evaluated, the difference
equations 
are solved at every depth point using the modified Feautrier
procedure given in Appendix I of RHI and the electron-scattering emissivity is
evaluated from Eq.\ (\eqref{2.2.3.5}).

The electron-scattering emissivity and opacity are taken together with those
of the free-processes as forming the ``background opacity and emissivity''
$\chi_c$ and $\eta_c$ which appear in Eqs.\ (2.6) and (2.7) of RHII. These are
then updated after every iteration, i.e. are treated by lambda iteration. The
cost of this operation per iteration scales as the product of the numbers of
depth and frequency points, i.e. in the same way as integration of the
monochromatic transfer equations.  In practice the additional time required
for the calculation of the electron-scattering emissivity  is hardly
discernable. However, the additional frequency points needed to  resolve
spectral features on the order of the electron-scattering Doppler  width do
lead to moderate increases in the computation time.

For coherent electron scattering by lambda iteration one simply sets
$E_j=J_j$. In RHII, coherent electron scattering was treated by direct
integration of the monochromatic transfer equation in vector form. For a
variety of test cases the results for coherent scattering calculated from the
direct solution and by lambda iteration were found to be identical, thus
confirming the validity of lambda iteration for electron scattering while all
other processes are included in the ALI procedure. However, for the realistic
model atmospheres discussed below, the  convergence from LTE initial
populations is very slow, and the iterations are started from  populations
calculated by the direct method. The iterative solutions for coherent
scattering then converges to exactly the same results as the direct solution.

\subsection{Discussion of results}

\begin{figure}
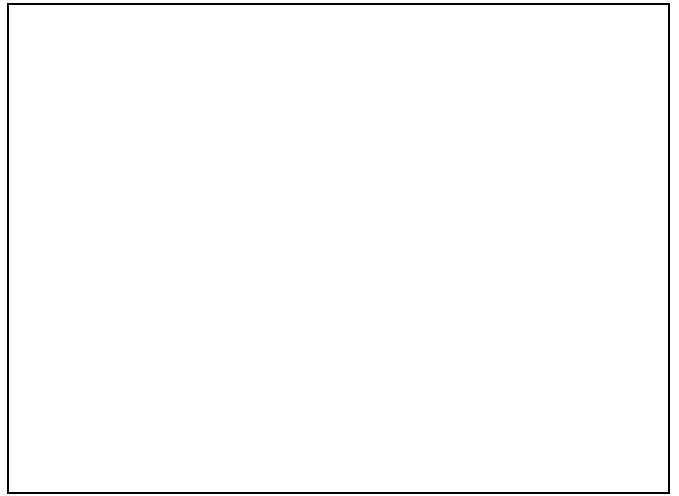

\picplace{6.5 cm}
\caption[]{Electron density and temperature as functions of mass column
density for Model 2}
\end{figure}

To illustrate the type of effects produced by noncoherent electron
scattering, we have input into MALI photospheric temperature and density
structures  from non-LTE models for early O-type stars, computed by Dietmar
Kunze. These models contain only hydrogen and helium; the relative abundances, 
effective temperatures and surface gravities are given in Table 2.  The  model
at $\Teff=50000\,{\rm K}$ is represented in Fig.\ 3; 
the other two are  qualitatively similiar.

\begin{table}
\caption[]{Model atmosphere parameters}
\begin{flushleft}
\begin{tabular}{llll}
\hline
Model & $\Teff$ &  $\log g$
    & $N$(He)/$N$(H)  \\
\hline
1&        40000\,K&       4.00&       0.100 \\
2&        50000\,K&       3.74&       0.100 \\
3&        60000\,K&       4.20&       0.111 \\
\hline
\end{tabular}
\end{flushleft}
\end{table}

In our calculations the temperature and  total atom density as a function of
the mass column density were specified and  held fixed. The original run of
electron  density was used only to calculate the initial LTE populations, and
subsequently the electron density was calcuated from the ion populations after
every  iteration. The converged electron density was found to be essentially
identical to that of the original  model, as were the level populations deep
in the atmosphere where the diffusion  approximation is valid. 
As distinct from the original models, the results obtained here for
noncoherent scattering will not be in strict radiative equilibrium
because of the changes in atomic level population.

We employed  simplified atomic models: \hneut\
 with five bound levels,
\heneut\ with 17 and \heplus\ with three. Doppler line profiles, 
which were used in the
calculation of the atmospheric structure, are also assumed here. This is not
realistic for the calculation of the surface flux, but permits the comparison
of coherent and noncoherent scattering without introducing further
complications, and should not influence our conclusions.

\begin{figure}
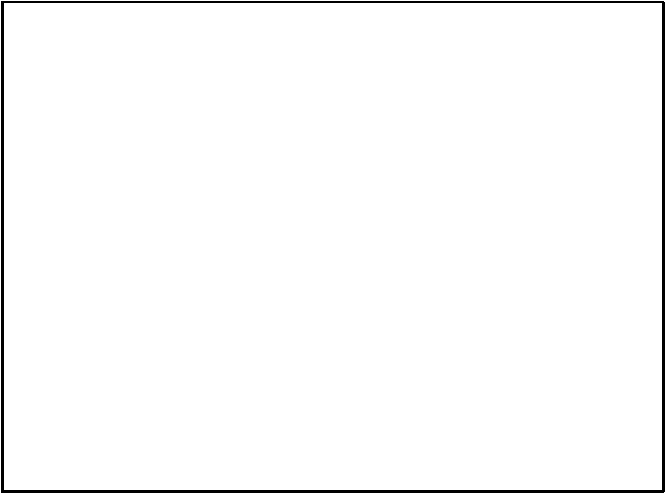

\picplace{6.5 cm}
\caption[]{Surface flux versus frequency for Model 2 computed with coherent
(dotted) and noncoherent (solid) scattering}
\end{figure}

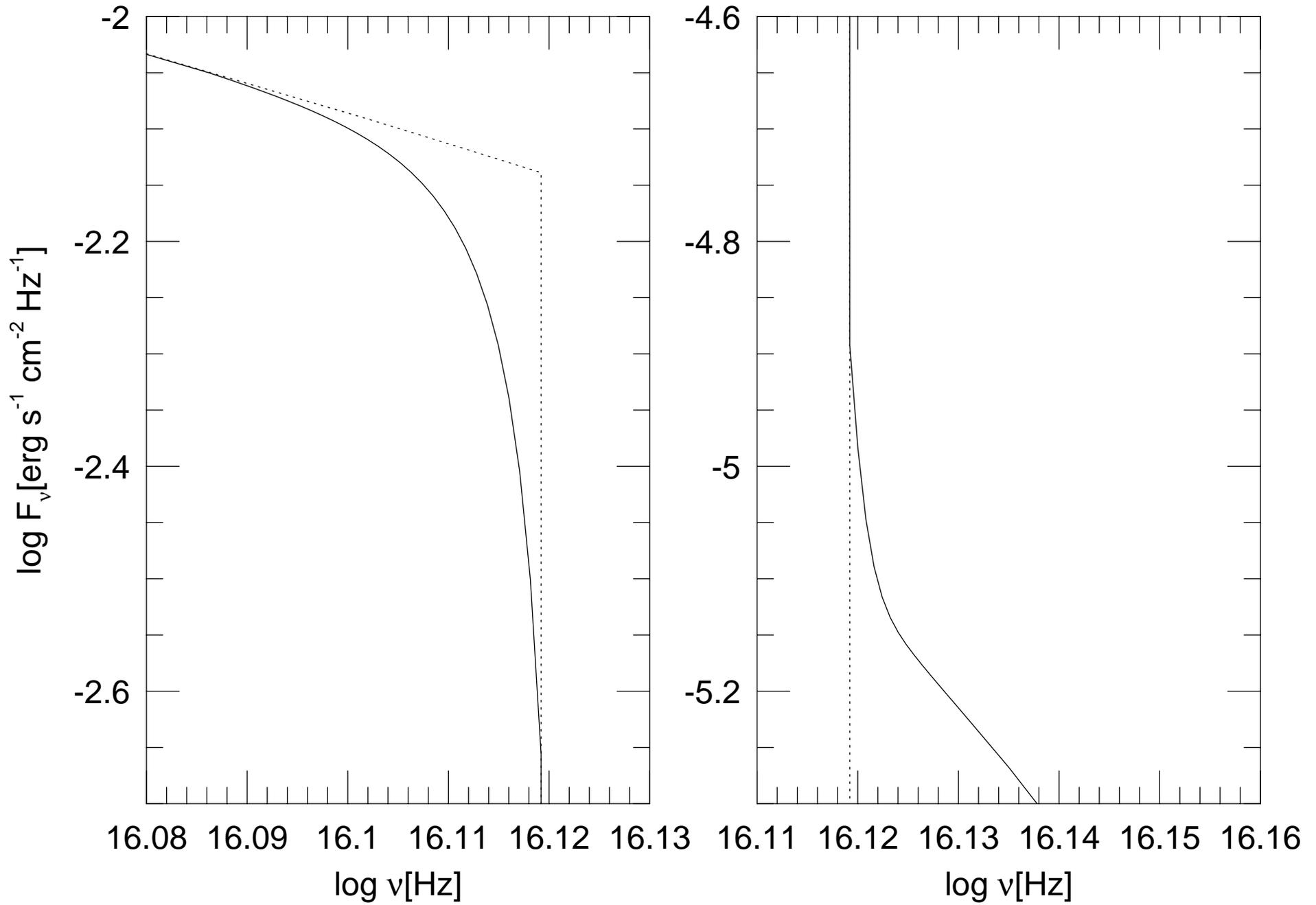
\begin{figure}
\picplace{6.5 cm}
\caption[]{Detail from Fig. 4 of the \heii\ continuum}
\end{figure}

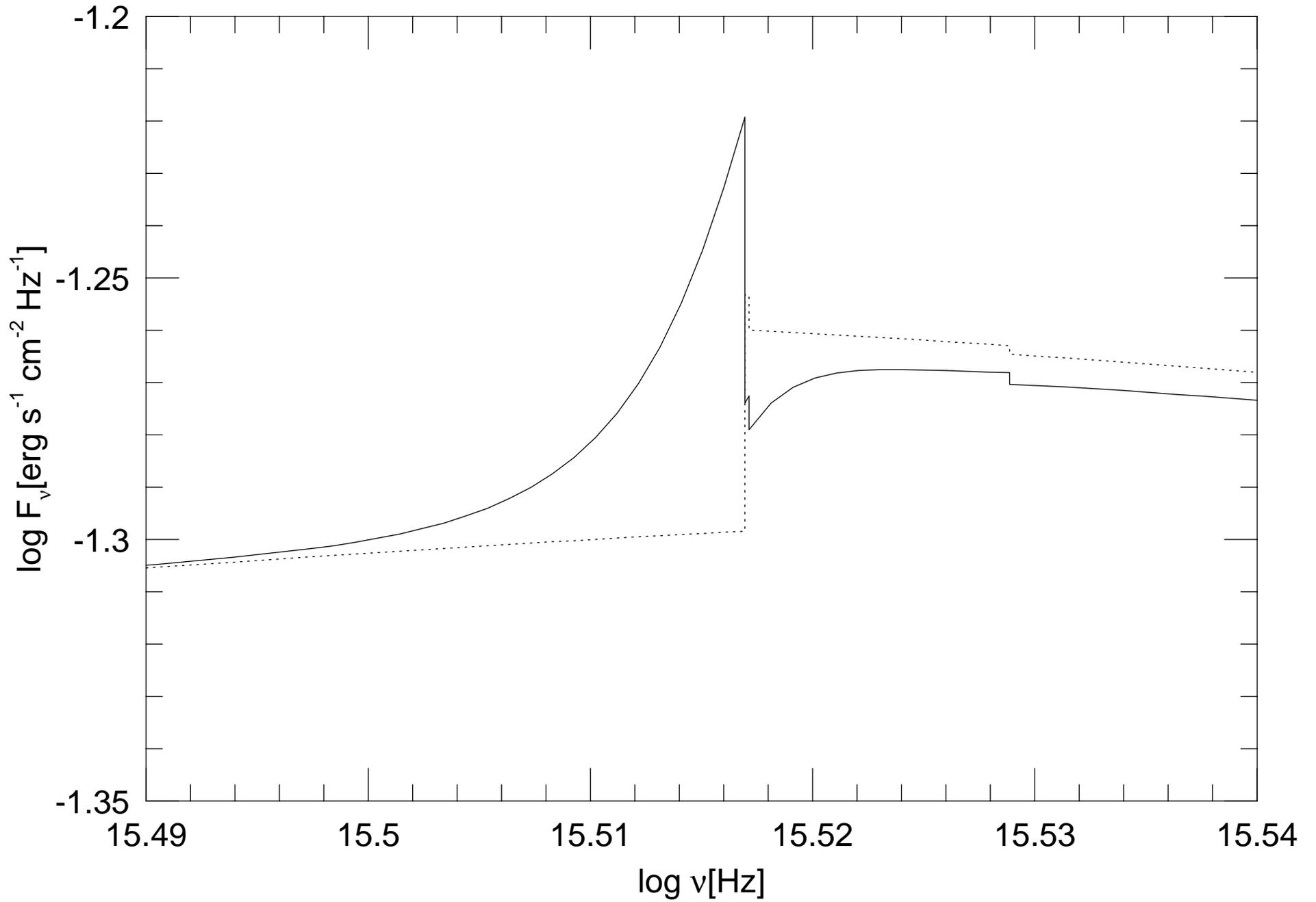
\begin{figure}
\picplace{6.5 cm}
\caption[]{Detail from Fig. 4 of the \hi\ continuum}
\end{figure}

\begin{figure}
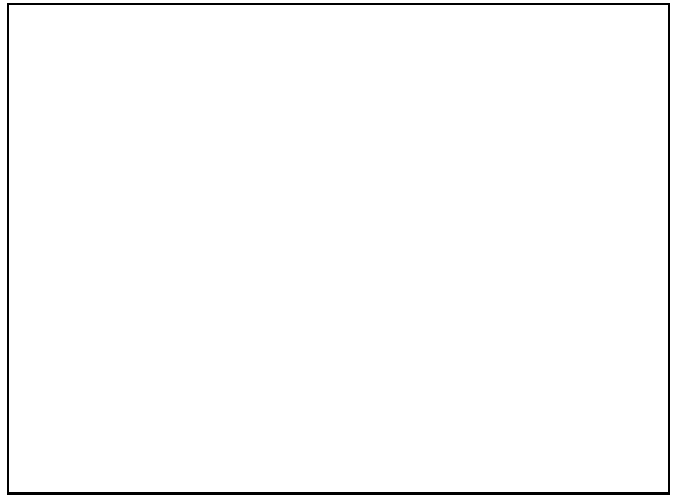

\picplace{6.5 cm}
\caption[]{Detail from Fig. 4 of the Lyman $\alpha$ line}
\end{figure}

\begin{figure}
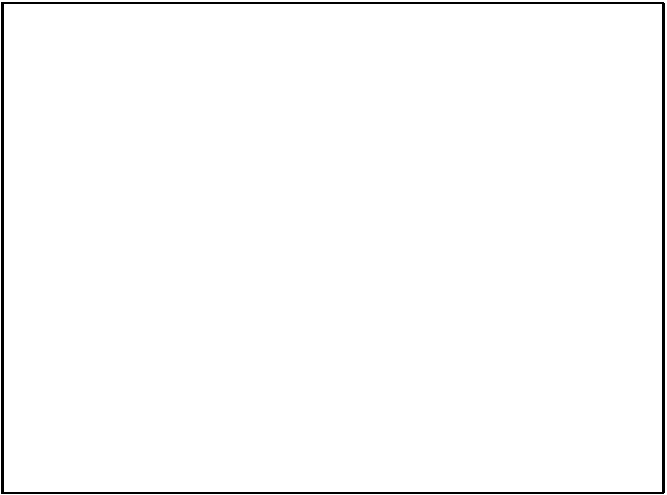

\picplace{6.5 cm}
\caption[]{Ionization fractions as functions of mass column density
for Model 2 computed with coherent
(dotted) and noncoherent (solid) scattering}
\end{figure}

We concentrate now on Model 2, which with $\Teff=50000\,{\rm K}$ 
and $\log g=3.74$
lies close to the Eddington limit.
The surface flux is shown in Fig. 4. The largest differences between the
results for  coherent and noncoherent electron scattering are  found in the
regions around the \heii\ edge, the \hi\ edge and  Lyman $\alpha$, which are
expanded in Figs.\ 5, 6, and 7, respectively. The  gas is essentially completely
ionized; the ionization fractions H$^0$/H, He$^0$/He and He$^+$/He are given in 
Fig.\ 8.  In Figs.\ 4--8 dotted and solid lines represent coherent and
noncoherent scattering, respectively.

Most conspicuous in Figs.\ 4 and 5 are the order of magnitude increase in the
flux in the \heii\ continuum and the decrease in the flux just below the  edge
caused by the noncoherent electron scattering. These effects arise from the
scattering by electrons of the intense radiation just below  the edge into 
the threshold region where the photoionization cross section has its maximum.
This increases the ionization rate of He$^+$ by an order of magnitude, and thus
shifts the ionization balance of He$^0$ by the same amount. The resulting
decrease in the populations of He$^0$ and He$^+$ are clearly seen in Fig.\ 8. 
Note that the additional flux in the \heii\ continuum arises 
not just from the photons scattering across the edge, but rather from those
escaping from the atmosphere because the the \heplus\ population is reduced.

The behavior at the \hi\ edge, shown in Fig.\ 6, 
is completely different. Now the only effect
of noncoherent electron scattering is appearance of the ``spur'' just below the
edge. This feature is caused by scattering of radiation from the Lyman 
continuum, which throughout much of the atmosphere is strongly in emission, 
to the region just below the threshold. This occurs even in models for which the
surface flux in the Lyman continuum is in absorption.
As no additional radiation is fed into the threshold region of the
Lyman continuum, and the amount lost is very small, the ionization balance 
is not significantly affected.

The narrow emission wings of Lyman $\alpha$ shown in Fig.\ 7 for coherent 
electron scattering are examples of
the Schuster effect. Non-coherence reduces this effect by scattering the
radiation into the very long wings. The slight deepening of the absorption
feature arises because the radiation trapped in the line core is scatterd out
into the wings. However, this effect is weak and in other cases the core is
slightly shallower with non-coherence. 

The general behavior of some of the
other spectral lines can be summarized as follows.
The Lyman $\beta$ line is similiar in appearance to Lyman $\alpha$ but less
deep. The three \hei\ resonance lines and \heii\ Lyman $\alpha$ are entirely 
in absorption with broad, very shallow wings. Balmer $\alpha$ is weakly in 
absorption, as are most of the He subordinate lines, although a few line of
\hei\ are weakly in emission with long shallow wings. All emission and 
absorption lines are weaker for noncoherent than for coherent electron 
scattering.

The above effects are found to a reduced extent in the model with  $\Teff =
60000\,{\rm K}$, $\log g=4.2$, and are present only weakly in the model  with
$\Teff=40000\,{\rm K}$, $\log g=4.0$. The ionization equilibrium in the surface
layers shifts by roughly factors of 3.5 and  1.7, respectively. This shows
that the decisive feature is the essentially complete ionization of  the gas, so
that the weak electron-scattering opacity is not completely overshadowed. 

It might also be objected that these phenomena will be radically reduced by the
lines converging on the series limit. Equating the electron Doppler width to
the ionization potential of a hydrogenic ion shows that the corresponding
principal quantum number is approximately
\be
n^* = 23.3\,T_4^{-1/4},  \e{aa}
\ee
where $T_4=T/(10^4\,{\rm K})$.
States with $n$ down to at least this value will be involved
in shifting of photons in or out of the continuum. Thus, two or three electron
Doppler widths at the temperatures considered here corresponds to a state 
below the region of confluence. In other words, radiation present below the
line merging region will continue to cause additional ionization.

\subsection{Effect of metal absorption}

The inclusion of metals is likely to reduce both the enhanced \heii\ emission 
and the \hi\ spur. However, for hot, low-gravity 
stars near the Eddington limit, absorption by {\em bound-free} 
transitions of metal ions should not substantially 
reduce the shift of the ionization balance, as species with 
appreciable abundances have ground state 
ionization edges lying in the \heii\ continuum, and thus will themselves
become more highly ionized, just as He itself. 

The effect of metal lines, on the other hand, is not easy to determine 
{\em a priori}. The dense array of metal lines (arising mostly from \feiv\ and 
\fev) between the \hi\ and \heii\ continua illustrated in Fig.\ 14c of 
Pauldrach et al (\cite{PK}) for a {\em plane parallel}
stellar model with $\Teff = 50500\,{\rm K}$ and
$\log g=3.785$ suggests that metal 
lines could reduce or quench entirely the additional ionization of He.
More appropriate would be {\em spherical} models including
wind effects -- the so-called unified models (Gabler et al 
(\cite{G}) and references therein). These give fluxes in the \heii\ continuum
larger than thoses in corresponding plane parallel models by some
orders of magnitude, which would shift the ionization equilibria of 
the metals to higher stages (the ionization potential of \feiv\ is only 0.4 eV
larger than that of \heii).  Moreover, deep in atmosphere of a hot star, the 
line blocking shifts to higher frequencies, while the scattering of 
radiation into the \heii\ continuum persists. The combined effects of these 
mechanisms remains to be investigated.

The main effects found here arise from the flow of radiation from deep hot 
parts of the atmosphere in relatively transparent parts of the spectrum, 
followed by scattering by electrons into the opaque regions, such as ionizing
continuum and strong lines.  Such effects are not, of course, limited to
stellar photospheres.  Winds of hot stars could be affected, as the scattering
of radiation in continuua would allow ions to receive ionizing radiation from
the photosphere which would otherwise be cut off by the increasing red-shift
of the stellar radiation as seen by the material in the wind. Clouds
illuminated by an external source of ionizing radiation could also experience
an increase in the degree of ionization by the same mechanism. Whether or not
these possibilities are realized in any particular case must be investigated. 
The technique developed here should make that possible.

\acknowledgements{We are indebted to Dietmar Kunze for the model atmospheres
used here, and to Keith Butler and Rolf Kudritzki for helpful discussions on 
the effect of metal lines.  We wish to thank the referee, Wolf-Rainer Hamann,
for useful comments on the paper. This work has been much facilitated by NATO 
Travel Grant 850674 to the Institute for Astronomy and Astrophysics of the 
University of Munich, and by the Smithsonian Institution Visitors Program.}

\appendix
\section{Appendix: The numerical method}

Applying the second-order Feautrier method to Eq.\ (\eqref{2.2.2.4}),
we obtain,
\bea
  -{\beta_T^2 \over b_i^2}
\left[  { F^{(i)}_{j-1}\over \Delta_{j-1/2}\Delta_{j}}\right.
  &-& \left. {2 F^{(i)}_{j}\over \Delta_{j-1/2}\Delta_{j+1/2}}
+{ F^{(i)}_{j+1}\over \Delta_{j+1/2}\Delta_{j}}\right]\nonumber\\
+  F^{(i)}_{j}
     &=&  J_{j},  \e{2.2.3.1}
\eea
for $j=2,\ldots,(\nf-1)$,
where, 
\bea
   \Delta_{j-1/2}&=&\xi_{j}-\xi_{j-1}, \qquad
   \Delta_{j+1/2}=\xi_{j+1}-\xi_j, \nonumber \\
   \Delta_{j}&=& {1 \over 2}(\xi_{j+1}-\xi_{j-1}).
   \e{2.2.3.2}
\eea

Equation (\eqref{2.2.3.1}) gives $(\nf-2)$ recurrence relations for the $\nf$
unknowns $F^{(i)}_j$.  In order to solve these equations, they
must be supplemented 
by appropriate boundary conditions at the ends of the grid.  
A particularly convenient set of boundary conditions  
are those of zero derivative,
which ensure the proper normalization condition (see 
Eq.\ [\eqref{2.2.2.5}] and the discussion following it).
The zero derivative conditions can be expressed to second-order accuracy
using the method of Auer (\cite{A}), which gives,
\bea
\left( {2\beta_T^2 \over b_i^2 \Delta ^2_{3/2}} +1 \right) F^{(i)}_1
  -{2\beta_T^2 \over b_i^2 \Delta ^2_{3/2}} F^{(i)}_2 &=& J_1 \nonumber\\
     -{2\beta_T^2 \over b_i^2 \Delta ^2_{\nf-1/2}} F^{(i)}_{\nf-1}
+\left({2\beta_T^2 \over b_i^2 \Delta ^2_{\nf-1/2}}+ 1\right)F^{(i)}_\nf
       &=& J_\nf  \e{2.2.3.6}
\eea

When applying these difference equations,
a problem arises at a continuum discontinuity,
where the radiation field $J$ has a separate left and right limit. 
In order to represent the radiation field near such a discontinuity
we assign two values for $J_j$,
namely, $J^-_j$ for the left limit and $J^+_j$  for the right. 
In this case, the value $J_j$ in 
Eq.\ (\eqref{2.2.3.1}) may be replaced with the quantity
\be
  {\tilde J}_j = {\Delta _{j-1/2} J^-_j + \Delta_{j+1/2} J^+_j
       \over \Delta _{j-1/2} + \Delta_{j+1/2}}, \e{2.2.5.1}
\ee
that is, with an appropriately weighted average of the
left and right limits of $J$ at the $j$th point.  It can be shown that
this procedure is equivalent to assigning
separate frequency points for the left and right limits and
then letting these two points approach each other (the proof of
this statement will be omitted here).

Equations (\eqref{2.2.3.1}) and (\eqref{2.2.3.6}) 
constitute a tridiagonal system of equations for the values of 
$F^{(i)}_j$ on the grid, which can be solved, as usual, 
by the method of Gaussian elimination. 
This is done
for each value of $i$ (i.e., two values for the $N$=2 exponential
approximations), and the results summed according to Eq.\
(\eqref{2.2.3.5}) to give the desired values of $E_j$.  
The operations count for this solution scales linearly with the
number of frequencies $\nf$.  Furthermore, the coefficient of
$\nf$ is quite small, since only simple algebraic operations are involved.
This compares very favorably with
other methods of applying a convolution operator to a very
unevenly spaced grid.

The preceding method for computing the emissivity
must be done separately for each spatial point (depth) in the
medium.  The coefficients of the difference equations 
(\eqref{2.2.3.1}) and (\eqref{2.2.3.6}) depend on depth, but
solely through the temperature dependence of $\beta_T$, so
a good deal of pretabulation of coefficients is possible.
The computation time necessary to find the emissivity at all depths
and frequencies scales simply as the number of frequency points 
times the number of depth points.
Typically these emissivity computations will represent only a moderate
fraction of the total time needed 
for the solution of the entire radiative transfer problem.  

\section{Appendix: Normalization of the method}

Using the discrete equations of Appendix A, it can be shown that
normalization condition (\eqref{2.11}) holds exactly, providing  
the integrals there are appropriately interpreted in terms of the
trapezoidal rule.  The derivation of this result will include the
possibility of discontinuities in $J$, as given in Eq.\ (\eqref{2.2.5.1}).
Starting with Eq.\ (\eqref{2.2.3.1})
(with $\tilde J_j$ replacing $J_j$), we multiply by $\Delta_j$ to obtain,
\bea
{1\over 2} F^{(i)}_{j}\Delta_{j-1/2} +{1\over 2}F^{(i)}_{j} \Delta_{j+1/2} =&&
{1\over 2} J^-_{j}\Delta_{j-1/2} +{1\over 2}J^+_{j}  \Delta_{j+1/2} \nonumber\\
   &&+(G^{(i)}_{j+1/2} - G^{(i)}_{j-1/2}), \e{b}
\eea
where $G^{(i)}_{j+1/2} \equiv (\beta_T / b_i)^2(F^{(i)}_{j+1}-F^{(i)}_{j})
/\Delta_{j+1/2}$.  Likewise, Eqs.\ (\eqref{2.2.3.6}) can be written,
\bea
{1\over 2} F^{(i)}_1 \Delta_{3/2} &=& {1\over 2} F^{(i)}_1 \Delta_{3/2}
   +G^{(i)}_{3/2} \nonumber\\
{1\over 2} F^{(i)}_{\nf}  \Delta_{\nf-1/2}&=& 
    {1\over 2} F^{(i)}_{\nf}\Delta_{\nf-1/2}  - G^{(i)}_{\nf-1/2}. \e{bb}
\eea
Now summing Eqs.\ (\eqref{b}) for $j=2,\ldots,(\nf-1)$, and adding in both Eqs.\
(\eqref{bb}), we obtain
\bea
\sum_{j=2}^{\nf} {1\over 2}F^{(i)}_{j} \Delta_{j-1/2}
   &+& \sum_{j=1}^{\nf-1}{1\over 2}F^{(i)}_{j} \Delta_{j+1/2}\nonumber\\
= \sum_{j=2}^{\nf}{1\over 2}J^-_{j} \Delta_{j-1/2}
   &+& \sum_{j=1}^{\nf-1}{1\over 2}J^+_{j} \Delta_{j+1/2}, \e{bbb}
\eea
Note that all of the $G^{(i)}$ terms have cancelled out.  Rearranging
summation indices, this can be written,
\be
  \sum_{j=1}^{\nf-1}{1\over 2} ( F^{(i)}_j + F^{(i)}_{j+1} )\Delta_{j+1/2}
= \sum_{j=1}^{\nf-1}{1\over 2} ( J^{+}_j + J^{-}_{j+1} ) \Delta_{j+1/2}
  \e{bbbb}
\ee
Multiplying by $a_i$ and summing over all $i$, using Eqs.\
(\eqref{2.2.1c}) and (\eqref{2.2.3.5}), we obtain,
\be
  \sum_{j=1}^{\nf-1}{1\over 2} ( E_j + E_{j+1} )\Delta_{j+1/2}
= \sum_{j=1}^{\nf-1}{1\over 2} ( J^{+}_j + J^{-}_{j+1} ) \Delta_{j+1/2}.
  \e{bbbbb}
\ee
This result is the discrete version of 
the normalization condition (\eqref{2.11}), where
the integrals have been
evaluated using the trapezoidal rule over each segment of 
the grid.

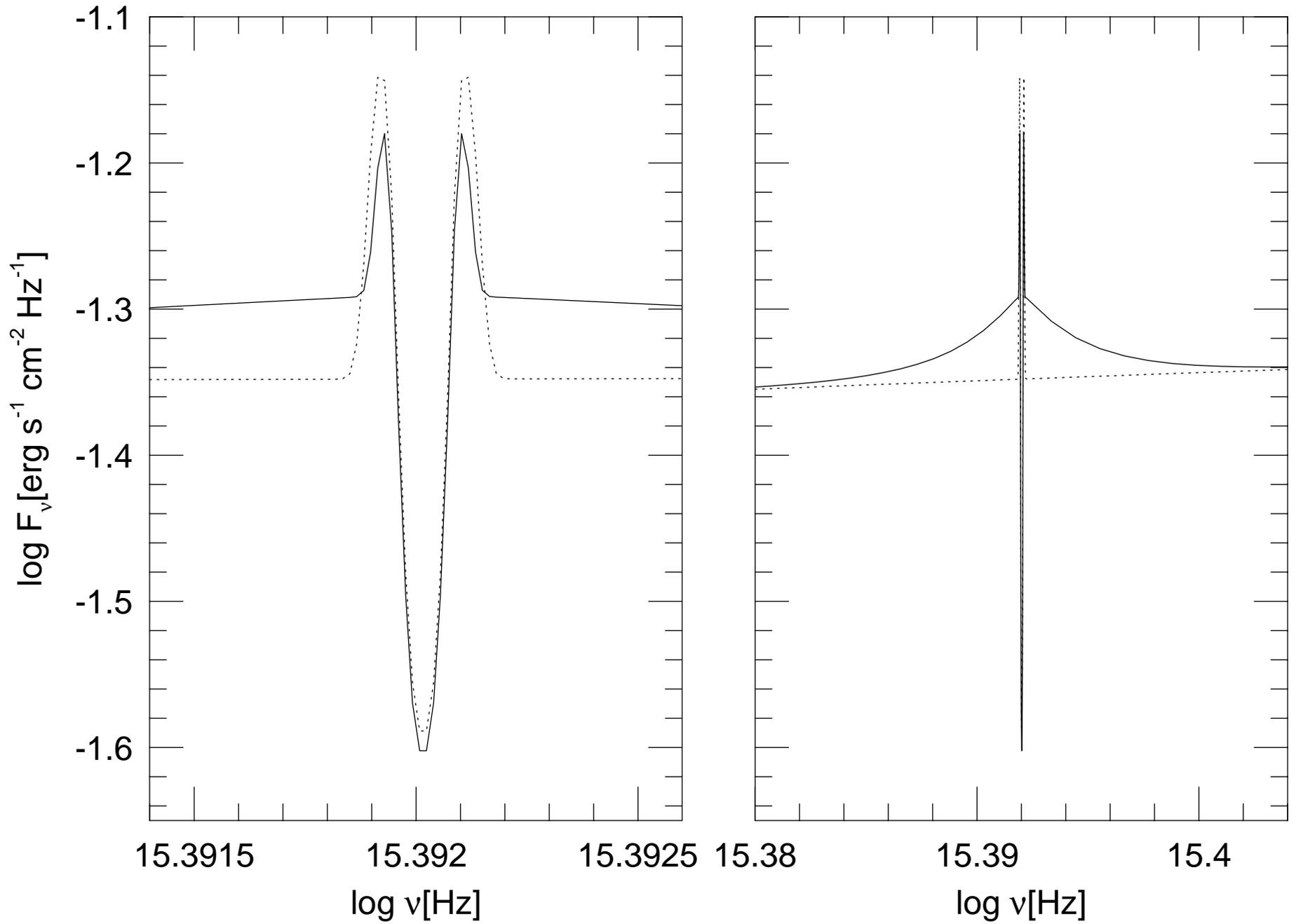
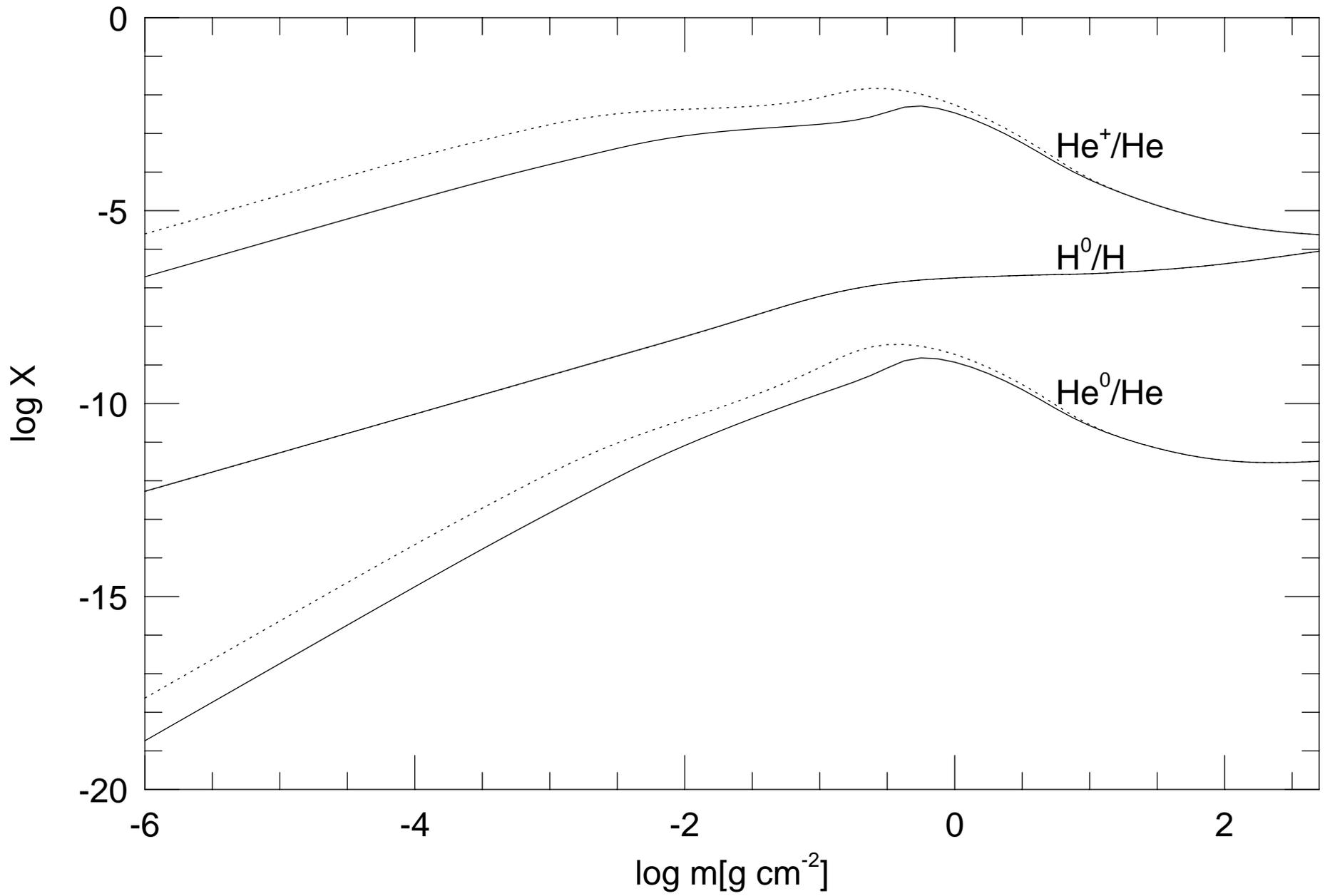
\end{document}